\documentstyle[12pt]{article}
\title{A direct pathway for sticking/desorption of H$_2$ on Si(100)}
\author{ P. ~Kratzer, B. ~Hammer and J. K. ~N{\o}rskov \\
 {\small \it Center for Atomic-scale Material Physics and} \\
 {\small \it Physics Department, Technical University of Denmark,} \\
 {\small \it DK - 2800 Lyngby, Denmark} }
\date{ }
\begin{document}
\thispagestyle{empty}
\maketitle
\begin{abstract}
The energetics of H$_2$ interacting with the Si(100) surface
is studied by means of {\em ab initio}
total energy calculations within the framework of density functional
theory.  We find a direct
desorption pathway from the mono-hydride phase which is compatible with
experimental activation energies and demonstrate the importance of
substrate relaxation for this process.
Both the transition state configuration and barrier height depend
crucially on the degree of buckling of the Si dimers on the Si(100)
surface. The adsorption barrier height on the clean surface is governed by the
buckling via its influence on the surface electronic structure.
We discuss the consequences of this coupling for adsorption
experiments and the relation between adsorption and desorption.
\end{abstract}
\section{Introduction}
In the last years, the H/Si(100) system has gained a growing attention
experimentally as well as theoretically. Apart from its technological
importance, this system is interesting because it is a simple
example for the interplay between hydrogen adsorption and the
reconstruction of a semiconductor surface.
Consequently, it has become one of the most intensively studied
adsorption system on a semiconductor surface.
The rich experimental material available for H$_2$ desorption from
Si(100) has raised a
number of questions,
and even quite fundamental issues are subject to controversy.
The remarkable discovery that this reaction obeys first order kinetics
\cite{SiSh89,HoLi92}
has started a lively debate over the desorption mechanism.
Among the suggested mechanisms, which included irreversible excitation
of hydrogen atoms into a band-like state \cite{SiSh89}, desorption mediated by
surface defects \cite{WuIo93,JiWh93a,NaSo94},
and concerted desorption from a ``pre-paired''
configuration \cite{DEYa92}, the last one has gained most support from
experimental observations.
While earlier experiments showed a considerable spread in the
activation energy for desorption, covering values of
$1.95$ eV \cite{SiSh90},
$2.17$ eV \cite{SiSh89},
$2.51$ eV and $2.86$ eV \cite{WiKo91}, more recent experiments agree
in an activation energy close to $2.5$ eV \cite{HoLi92,FlJo93}
for well-prepared, ordered surfaces.

The energy distribution within the reaction product H$_2$ has been
studied extensively in a number of experiments.
Roughly 1\% and 0.2\% of the H$_2$ molecules desorbing from the
mono-hydride and di-hydride phase, respectively, are in the first
vibrationally excited state. In both cases this is a factor of 20 more
than one would expect for molecules in equilibrium with the surface.
On the other hand, the rotational temperature of the molecules is
lower than the surface temperature \cite{ShKo92b}. From these findings
and measurements of the translational energy it has been concluded that
the total energy of the desorbing molecules shows no evidence for a
substantial barrier for adsorption \cite{KoNe94}.
On the other hand, sticking of H$_2$ on Si(100) could not
be detected until recently . This has raised speculations about a
high barrier towards dissociative adsorption, which seems to be difficult to
reconcile with the desorption data. Very recent experiments, however,
show that some sticking is possible at elevated surface
temperatures.
Kolasinski, Nessler, Bornscheuer and Hasselbrink have shown that sticking
from a molecular beam is activated
with respect to the total energy of the adsorbing molecules, but also with
respect to the substrate temperature\cite{KoNe94a}.
Bratu and H{\"o}fer found an even stronger activation of
sticking with surface temperature for a thermal gas of both H$_2$
and D$_2$ \cite{BrHo94a}.
These findings point towards the importance of the substrate degrees
of freedom for a possible solution of the barrier puzzle.
Structural considerations give further support to this picture:
For the clean Si(100) surface the existence of the `buckled dimer'
reconstruction has been established experimentally \cite{Wolk94}
as well as theoretically \cite{DaSc92}.
In the (2x1)-phase of the H-covered surface, however,
the dimers are symmetric, suggesting that
considerable rearrangement of the substrate atoms must be involved in the
adsorption process.

Earlier theoretical efforts have concentrated on total energy
calculations for Si clusters based on ab initio Hartree-Fock/CI
methods \cite{WuIo93,JiWh93a,NaJo91}.
The calculations of Wu, Ionova and Carter showed that the barrier
height is sensitive to substrate relaxation \cite{WuIo93}.
Under the assumption that H$_2$ desorption proceeds in a single
step, part of the results \cite{WuIo93,JiWh93a}
were at variance with the observed thermal activation energies.
Therefore, these two groups argued in favor of a two-step
desorption mechanism via SiH$_{2}$ groups formed at defect sites.
Such a mechanism, however, is compatible with first-order reaction
kinetics only if diffusion of hydrogen atoms on the surface is the
rate-limiting step, which is very unlikely in the light of recent
modeling of the diffusion process \cite{ViSe93,WuCa94}.
The agreement in the desorption rate and the reproducibility
of the activation energy in recent
experiments also argues against the role of defects.
Recently substantial progress has been made towards a better
understanding of the energetics of H$_2$ desorption.
By comparing two cluster calculations, Nachtigall and coworkers
\cite{NaSo93} showed that a description of exchange and correlation
based on non-local density functional theory gives
a  Si-H bond energy in fair agreement with, though systematically
lower than conventional quantum
chemistry methods. The heat of chemisorption derived from their results is
compatible with the observations, although their cluster calculations
give a too high desorption barrier for a direct desorption mechanism
to work \cite{NaSo94}. However, recent slab
calculations \cite{KrHa94,PeSc94,ViSe94}
agree in a fairly low adsorption barrier, and desorption through this
transition state gives a barrier in good agreement with the
experimental values.
Therefore, we see no further need to invoke a defect-mediated mechanism for
desorption on energetic grounds.

In this paper, we present {\em ab initio} calculations for
the energetics of H$_2$ adsorption on
a laterally infinite slab. A preliminary account of the present
calculations was given in ref.\cite{KrHa94}.
We discuss in detail the coupling of
adsorption to the substrate, making advantage of the fact that
the reconstruction of the clean Si(100) is well described in our approach.
We discuss the consequences of our findings for the dynamics of
desorption and adsorption and relate our results to the experimental data.

\section{Calculations}
We performed total energy pseudopotential calculations within a basis
set consisting of plane waves. In this approach Hellmann-Feynman
forces on the ions can easily be derived. These forces are used to
relax the ions and to determine the stable structures.
Unless otherwise stated, the slab
consisted of a periodically repeated 2x1 unit cell in the lateral
directions, build up from 6 layers of Si atoms in the z direction,
separated by a vacuum region corresponding to another 6 layers.
Test calculations were performed for an 8 layer slab as well.
Both of these slabs contained a glide plane
in the middle, with hydrogen adsorbing on both sides. To demonstrate
that our calculations are reasonably converged with respect to layer
thickness, we also tried an asymmetric configuration with the
adsorbing H$_2$ on one side of the slab, the dangling bonds on the other side
being saturated with (static)
di-hydride units. Results of these convergence tests are collected in
table 1.

Our approach is based on the density functional theory. For the
exchange and correlation energies we can employ either the local density
approximation (LDA) \cite{CeAl80} or the generalized gradient
approximation (GGA) \cite{PeWa92}.
For systems with H atoms in different chemical environments, such as
free H$_2$ compared to H adsorbed on surfaces, GGA
calculations have been shown to give reliable results, whereas the LDA
calculations are at variance with experimental data
\cite{HaJa93,GuJa94,HaSc94}.
Similar observations have been reported for CO adsorption on metal
surfaces \cite{PhVe94,Ki??94}.
This leads us to the conclusion that the GGA is the method of choice for
our system.
We have calculated electronic densities self-consistently in the LDA,
which serve as an input for the non-local exchange-correlation
expression of the GGA. This approximation has been shown to work
accurately for a number of systems \cite{HaJa93}.

The electronic ground state of the system is found by conjugate
gradient minimizations of the total energy \cite{PaTe92},
interlaced by subspace rotations as suggested by Gillan \cite{Gill89}.
Since the unreconstructed Si(100) surface is
metallic, it is necessary to treat
partially occupied bands. In order to stabilize the process of
determining the electronic ground state, we introduced a finite
electronic temperature of $k_B T=10$meV as a technical parameter,
which allows us to obtain occupation numbers from a smooth Fermi
distribution. The smallness of the smearing assures
that the semiconducting
character of bulk silicon is unaffected. Quoted electronic energies
have been extrapolated to zero electronic temperature.

To model the Si atoms we employ the sp-nonlocal pseudopotential of
Bachelet, Hamann and Schl{\"u}ter \cite{BaHa82}, which was brought to the
Kleinmann-Bylander form \cite{KlBy82}. Hydrogen is described by the Coulomb
potential of a proton. The numbers quoted in this paper were obtained
using a plane wave basis set with kinetic energies up to 30 Ry. This
basis set quality is mainly dictated by the requirement to represent
accurately the electronic wavefunctions and binding energies of
hydrogen.
Tests were done also at 40 Ry (see table 1). K-space sampling was
performed with a density corresponding to 32 k-points in the first
surface Brillouin zone (SBZ) of the 2x1 unit cell. The symmetry of the slab
was exploited to reduce the number of independent k-points actually
used in the calculations \cite{Cunn73}.
Computing the bulk silicon properties under these conditions, we
obtained a bulk modulus of
0.97 MBar and a lattice constant of 5.394 \AA, which was used
throughout the further calculations.

During the search for the transition state both the hydrogen atoms
and the Si atoms of the two outer
layers of the slab were allowed to move within the plane containing the dimer
bond. At least two inner layers of the slab were kept static.
Since the transition state is a local maximum in the direction of the
reaction coordinate, it is necessary to impose some constraint (e.g.
fixed H-H distance)
to prevent the system from sliding down the reactant or product valley.
Calculations were repeated
with different constraints until the transition
state could be pinned down. The reaction path was then determined by
unconstrained steepest descent minimization on both sides of the
saddle point. For the sake
of computational feasibility this part of the calculation was
performed with a 10 Ry basis set cut-off and a sampling
quality of 8 k-points in the 2x1 SBZ only.
However, we stress that all energies quoted and plotted in this paper
are based on the more converged parameter set (30Ryd and 32 k-points).

\section{Results}
\subsection{Surface Structure}
As a starting point we determined the stable structures of the
H/Si(100) system by relaxing the adsorbate and the two Si layers
beneath it. For the mono-hydride in the (2x1) phase we obtain an
equilibrium geometry with a distance of 2.44{\AA} in the Si dimer.
The Si--H bond is 1.51{\AA} long and forms an angle of $112.3^o$
with the Si dimer axis.
The clean surface consists of buckled Si dimers with a bond length
of 2.28{\AA}.
In the following, all energies refer to
the sum of the energy of the H$_2$ and the clean, buckled  surface.
On this scale, the hydrogenated Si dimer has a total energy of $-2.14$eV, which
corresponds to a mean Si--H bond strength of 3.45eV.
The isomeric structure with a di-hydride at one Si atom and two
dangling bonds at the other is metastable and
has an energy of -0.45eV on this scale. The
difference is mainly due to the absence of the Si dimer bond, while
the Si--H bond strength only decreases slightly to 3.32eV. For the length
of this bond we obtained 1.55{\AA}.

In the following we restrict ourselves to the 2x1 mono-hydride phase, which is
the ground state of the system for low H coverage. Since the clean Si
dimers have additional $\pi$-bonding, it is energetically
more favorable for two H atoms to sit at the same Si dimer than at
two different ones. An estimate for the energy gain derived from
experiment gives 0.25eV \cite{HoLi92}, the best theoretical estimate at
the moment is  about 0.15eV \cite{NaSo93}. Due to this effect
the hydrogen atoms are ``pre-paired'' prior to desorption
for coverages that are not too low. The
desorption rate is therefore proportional to the coverage, rather than
the coverage squared, as usually expected for associative desorption.
Restricting our further
investigation to the desorption from such a ``doubly-occupied'' Si
dimer, our mechanism will be  a first order process, in accordance
with experimental evidence.  We shall first
concentrate on the minimum energy pathway for desorption, which is
highly asymmetric, and then discuss the alternative of a symmetric
pathway.

\subsection{Desorption pathways}

Fig. \ref{adiabat} shows the calculated transition state and minimum energy
path assuming that the substrate relaxes completely
during each step of the reaction (i.e. the adiabatic limit of
substrate motion).
As a first step in desorption from the mono-hydride phase one
hydrogen atom
moves over to the other side of the Si dimer. Simultaneously the
dimer itself changes from a symmetric to a buckled configuration.
The two H atoms pass through a highly asymmetric transition
state. Finally the H$_2$ molecule leaves the
surface, while the Si dimer buckles even stronger until it has reached its
equilibrium configuration on the clean surface.
The details of the transition state are given in fig.\ref{trasta}(a).
Our findings give evidence for a strong coupling of the desorption
process to the Si dimer buckling angle, while the Si-Si distance within
the dimer at the transition state is almost the same as in the mono-hydride.
Since the final step essentially
takes place at a single silicon atom, our mechanism fits well
into the picture of a
common exit channel for H$_2$ desorption from all silicon surfaces,
which has been
invoked to explain similar final state distributions for H$_2$ from
different surface orientations and adsorbate phases observed in
experiment \cite{ShKo92b}.
We would like to point out, however, that the system never passes
through the metastable di-hydride species mentioned above, but
follows a direct route from the adiabatic transition state to the
mono-hydride, as can be seen from the energy diagram in fig.\ref{energy}.

If we require the mirror symmetry of the mono-hydride phase to be
retained during desorption, we arrive at the symmetric transition
state shown in fig.\ref{trasta}(b).
The breaking of the H$_2$ bond occurs already with the molecule further
out above the surface as compared to the asymmetric pathway.
Since it is only 0.08 eV higher in energy than the asymmetric one,
contributions to the desorption rate
cannot be excluded from simple activation energy arguments.
Instead, a more
detailed kinetic analysis is needed, which accounts for the different
prefactors for each pathway. We have
therefore included the symmetric transition state in the following analysis.

The desorption from the Si(100) surface is comparable to H$_2$
abstraction from disilane, where the
Woodward-Hoffmann rules have been invoked to explain why an
asymmetric transition state is favored \cite{GoTr86}.
There are, however, important differences between the reverse
processes, the adsorption and the addition of H$_2$ to disilene.
The latter process has a symmetric transition state, and  the
$\pi$-bond in the
surface Si dimer is much weaker due to geometrical constraints on the
surface, compared to $\pi$-bonding in disilene.
For both reasons we get a substantially lower adsorption barrier compared
to the barrier of 1.65eV for (symmetric)
addition of H$_2$ to disilene \cite{GoTr86}.

\subsection{Coupling to the substrate}

Because of the strong coupling of the dimer motion to desorption we also
expect a strong dependence of the barrier for sticking on the
substrate configuration. There is, however, an important difference
between sticking and desorption.
In a thermal desorption experiment, the system probes a large part of
the multi-dimensional potential energy surface before it (most likely)
finds its way over the lowest transition state. This situation should
adequately be described by a kinetic theory including all relevant
degrees of freedom.
In an adsorption experiment, the molecule will either overcome the
barrier towards sticking or immediately be reflected to the gas
phase.
In a first approximation, we can assume that each H$_2$ molecule
probes a single frozen substrate
configuration in its encounter with the surface. Most of these
configurations  will exhibit a large adsorption barrier, but
sometimes a thermal fluctuation
of the surface will create a configuration favorable for sticking.
The observed sticking could be completely determined by these thermally
activated events. Such a scenario has been proposed to
resolve the apparent contradiction between
desorption and sticking experiments \cite{KoNe94a,BrGr94}.

In this context the question arises which modes of the silicon dimer
are the most important ones for such an activated sticking.
Fig. \ref{adiabat} clearly shows that the lower Si atom of the dimer
changes its position, while the upper one moves only little. Apart
from a minor change in the Si dimer bond length, this motion can be
described as a change of the buckling angle. We therefore investigated
the influence of the buckling on the asymmetric transition state.
Additionally, we have
studied the importance of the Si dimer stretch mode for the symmetric
transition state.

To order to explore the coupling between the H$_2$ molecule and the Si dimer
buckling in more detail, we tried to place H$_2$ in a
transition state geometry asymmetrically on both a fully buckled dimer
and a symmetric dimer.
In fig. \ref{barriers} the total energies $E_{TS}(x)$
of these configurations
(relative to the equilibrium surface energy plus the energy of the
free molecule) are shown as a function of buckling angle $x$.
The lower curve, $E_{s}(x)$, displays the energy of the
corresponding Si dimer configuration on the clean surface.
Its minimum is the result of a balance between two effects:
The buckling increases the splitting between the bonding and
antibonding state derived from the two dangling bonds in the Si dimer,
until a gap opens and the surface becomes semiconducting.
On the other hand, it
introduces additional surface strain, which limits the favorable
amount of buckling.

Within the simple static picture outlined above, we can introduce the
difference $E_a(x)=E_{TS}(x)-E_s(x)$ as the barrier height felt by a
molecule hitting a surface frozen in that particular configuration.
Alternatively, it can be interpreted as the energy gained by a
desorbing molecule after traversing a saddle point with the substrate
coordinates held fixed.

A similar analysis for the coupling to the stretch
mode of the symmetric Si dimer is made in fig.\ref{stretch}. We
continuously weakened the Si dimer bond by enforcing a fixed distance $y$ in
the Si dimer and determined the symmetric transition state for each of these
configurations. For a reasonable Si dimer stretch we find only minor
changes in the static barrier $E_a(y)$. Only for the unrelaxed, unpaired
Si(100) surface this barrier decreases to 0.15eV.
Although unpaired
silicon atoms occur only very rarely on a well-prepared Si(100)
surface, this result indicates that the barrier for sticking can
become unusually low in this situation.
We expect a similar conclusion to hold for single atom defects.

We also tested the stability of the symmetric transition state against
buckling of the Si dimer (dashed line in fig. \ref{barriers}).
With increasing buckling the most favorable transition state
is seen to move over from the symmetric configuration to the
lower silicon atom of the dimer.
To explain this tendency it is beneficial to consider the energetics of
adsorption as governed by a balance between repulsive effects from the
filled orbitals of the adsorbing molecule and attraction caused by
hybridization of the molecule's antibonding states with the electronic
states of the substrate \cite{LuGu79}.
The energy of
the dangling bond state of the lower/upper silicon atom is raised/lowered in
energy during the buckling \cite{ZhSh89}, which results in electronic
charge being transfered from the lower to the upper Si atom. This
effect is clearly visible in fig.\ref{chd}. The H$_2$
molecule can minimize its repulsion with the surface
by approaching the depleted lower end of the Si dimer.
For that reason, buckling tends to facilitate dissociation.
On the symmetric Si dimer,
both dangling bonds are occupied with one electron each.
Placing the H$_2$ above one end of the symmetric Si dimer causes
repulsion between the electrons
in the $1\sigma_g$ state of hydrogen and the electron in
the dangling bond state, and is energetically unfavorable
(cf. fig. \ref{chd} and \ref{barriers}).
In this situation, the symmetric pathway is prefered.
A discussion of the attractive interaction between the molecule and
the surface partially counterbalances the above trends. As can be seen
from fig.\ref{surfbands}, the splitting of the bonding and
antibonding surface states is increased by the interaction with the
H$_2$ molecule during adsorption and the antibonding state is
partially removed from the gap. The energy gain is largest when the
antibonding state was yet partially occupied before the interaction,
i.e. when the buckling was too small to make the clean surface
semiconducting. As a result of both the repulsive and the attractive
interactions, the lowest adsorption barrier occurs for some
intermediate value of the buckling angle.

\subsection{The transition states}
 The observed very low sticking has raised the question of an entropic
contribution to the adsorption barrier. To quantify the steric
constraints for the reaction,
we performed a normal mode analysis for the molecule at both the
lowest asymmetric and symmetric transition state.
The results are shown in table 2.
For the asymmetric transition state, the four degrees of freedom
of the molecule moving in the plane of the Si dimer are found to be
strongly coupled.
The reaction coordinate has components both in the direction of
the stretch of the molecular bond and
the motion of the molecule's center of mass parallel to the
surface. This reflects the observation that one hydrogen is stripped
from the molecule as it passes the lower silicon atom.
The frequencies of the real modes are fairly high compared to
metallic systems \cite{HaSc94,HaJa92,WhBi94},
indicating strong configurational constraints for
the reaction to take place. This can be rationalized by the covalent
character of the bonding on the Si surface, which leads to a strong
corrugation of the potential. The sum of all real frequencies, 585meV,
even exceeds the vibrational energy quantum of H$_2$.

The symmetric transition state lies further out above the surface. Here
the molecule feels a less strongly corrugated potential, which is
reflected by the low lying modes associated with the variation of
the molecule's impact position within the unit cell. The mode
associated with the bond stretch, however, has only dropped to
one half of its gas phase value.
Again, the sum of all real modes, 475 meV, is sizeable.

As a consequence, the barrier for adsorption through both the
symmetric and the asymmetric transition
state depends extremely strongly on the impact
parameter in the surface unit cell and the orientation of the incoming
molecule. Only a tiny fraction of the molecules that have just enough
energy to surmount the minimum energy barrier will actually impact the
surface with the right configuration. As the energy of the incoming
molecule increases, the "hole" in configuration space where it
can get over the barrier will increase, thus leading to a slow rise of
the sticking coefficient.

\section{Discussion}
\subsection{Desorption}
Combining the obtained information about the potential energy surface
of the H$_2$/Si system
with some simple dynamical considerations can shed some new light on
the puzzling experimental data.

To check the reliability of our results, we first give
an estimate for the activation energy for desorption which includes
zero-point energies. We have calculated the vibrational
frequencies of the mono-hydride to be 256 meV for the Si-H stretch
(the experimental value is 260meV, see ref. \cite{TuCh85})
and 77 meV for each of the bending modes.
Including all zero-point energies at the transition state, we get a
desorption energy of 2.50 eV (see fig. \ref{energy}),
in excellent agreement with the
most reliable experimental values \cite{HoLi92,WiKo91,FlJo93}.

We can also account for the observed vibrational excitation of the
desorbing molecules.
At the calculated transition state the H-H bond is stretched by 35\%
and the associated frequency is lowered (see fig. \ref{PES}). Together
with the potential drop after the barrier this will give rise to molecules
desorbing vibrationally excited \cite{KuBr91}, in qualitative agreement with
the observations \cite{ShKo92b}.

To comment on the energy in the other degrees of freedom of the
molecule a more careful dynamical treatment is needed. Because of the
confined transition state and stiff normal modes, transition state theory
would predict that a sizeable amount of zero point energy should end
up as kinetic energy of the molecule. As mentioned earlier, the sum of
the normal mode energies is of the order of the H$_2$ vibrational
quantum for both transition states we have examined, and therefore the
{\em total} excess energy of the desorbing molecules is essentially
given by $E_a(x)$ in this model. However, transition state theory may not
be applicable, because desorption is too fast to establish thermal
equilibrium at the transition state, and only a fraction of the zero
point energy will actually contribute to the final energy of H$_2$.
Because of the strong mixing of the molecular modes at the transition
state the detailed energy distribution among the vibrational,
rotational and translational degrees of freedom will have to await
quite elaborate calculations of the desorption dynamics.

It should be mentioned that the surface phonons will also take up part
of the energy released after passing the desorption barrier. Since
the surface atoms are considerably heavier than hydrogen, it is a reasonable
first approximation to assume that they will stay in the saddle point
configuration during desorption. In this limit the energy released
to the substrate coordinates can be read from the $E_s(x)$ curve
in fig.\ref{barriers}
and amounts to 0.09eV for the most favorable configuration.
In a more realistic treatment, the desorbing H$_2$ molecule will
transfer momentum to the substrate, thereby causing the Si dimer atoms
to recoil. We therefore expect the above number to be a lower limit
for the energy transfer to phonons.
A detailed investigation of this effects is currently in
progress \cite{LuKr95}.

Nevertheless, our calculations suggest that a sizeable amount of
energy, up to 0.4eV, depending on the desorption pathway,
the exact barrier height and the role of zero point energy, will end
up in the molecular degrees of freedom. Experimentally, it has been
found that the molecules have $0.077\pm0.08$eV in excess of their
thermal energy after desorption \cite{KoNe94a}.
Although a quantitative comparison of
theory and experiment will have to await a full dynamical treatment,
this finding seems to indicate that our GGA calculation overestimates
the barrier slightly. We tried different slab geometries and basis-set
sizes
and found an uncertainty in the barrier height of the order of
0.1eV (see table 1).
Further sources of systematic errors could be  the interaction between
hydrogen in different unit cells and the non-selfconsistent treatment
of the GGA corrections in our approach, and last, but not least the
GGA itself.  Indeed, similar calculation
using a larger unit cell have reported a somewhat lower barrier
height \cite{PeSc94}.

\subsection{Adsorption}
The apparent absence of sticking on the cold clean surface is readily
explained by the static barrier height of 0.67eV together with
the strong steric constraints on the transition state we have found.
Heating the surface will excite the Si dimer to a less buckled
configuration with a lower barrier and thus promote sticking.
Both effects have been demonstrated in a simple dynamical
model, which treats the slow degrees of freedom in a classical sudden
approximation \cite{KrHa94}.
The modulation of the barrier height with buckling angle results in a
sticking coefficient increasing with surface temperature with
an apparent activation energy of 0.09 eV for beam energies up to
about 0.3 eV, and slightly lower at higher energies.
A more elaborate treatment of the dynamics, taking into account two
molecular and one substrate degree of freedom, gave similar results
 \cite{RuBr95}.
As has been pointed out before \cite{KrHa94},
this  activation energy can simply be interpreted as the energy
$E_s(x_m)$ required
to bring the clean surface Si dimer to the angle $x_m$ where $E_a(x)$ has
its minimum. At low kinetic energies the adsorption process is therefore
dominated by the minimum barrier $E_a(x_m)$, and the rate
is directly proportional to the probability that a thermal
fluctuation brings the surface in a configuration of a lower barrier
for the reaction.
In a realistic model of sticking, the dynamical coupling between
the impinging molecule and the substrate has to be taken into account
as well.
Although such a coupling seems to be suppressed by the large
mass mismatch between
hydrogen and silicon, it will certainly effect the sticking probability
for incident energies substantially lower than the barrier height,
when tunneling is important.
The so-called ``thermally assisted tunneling'' can give rise to a
strong increase of the sticking coefficient with surface temperature,
and when being
interpreted as an apparent activation energy, its value will vary
strongly with the incident energy of the molecules \cite{LuHa91}.
The implications of this coupling for the sticking of D$_2$ on Si(100)
are currently being investigated \cite{LuKr95}.

The angular dependence of the sticking has not been measured yet.
{}From the normal mode analysis of this transition state we find that
the reaction coordinate couples strongly to the motion of the molecule
parallel to the surface. We therefore conclude that parallel momentum
is probably important in accessing the transition state. Hence we
expect the sticking probabilities
not to obey normal energy scaling as a function of incident angle.

\section{Conclusion}
In conclusion, we have shown that {\em ab initio} calculations give
evidence for a decisive role of surface structure in the
interaction of H$_2$ with the Si(100) surface. We find a
direct desorption pathway from the mono-hydride phase in accordance with the
observed activation energy. Combined with the pre-pairing mechanism,
this also explains the peculiarities of reaction order on this surface.
Adsorption is strongly hindered both by a
sizeable static barrier and a strongly confined transition state.
Both the barrier height and location couple strongly to the surface
modes, and we find a buckling mode of the silicon dimer to promote
the sticking very efficiently.

Finally, we would like to point out that this is the rationale
behind both the phonon-assisted sticking and the apparent
lack of ``detailed balance'' in the experiments.
Although we find a unique lowest transition state,
adsorption and desorption experiments are not reciprocal to each
other, as is usually assumed in the application of ``detailed
balance'' arguments, since the initial and final configurations of the
silicon dimer are different
in both adsorption and desorption (actually they are interchanged).
As a consequence, adsorption and desorption
experiments reveal different aspects of the H$_2$/Si(100) interaction
potential.

\section{Acknowledgments}
The Center for Atomic-scale Materials Physics is sponsored by
The Danish National Research Foundation. Our research was supported
by The Danish research councils through the Danish Center for Surface
Reactivity. P.K. acknowledges a grant from
the EEC Human Capital Programme ``Energy Pathways in Bond Making and
Breaking at Surfaces'' (CHRX-CT93-0104).

\renewcommand{\baselinestretch}{1.0}

\newpage
\begin{table}[h] \label{conver}
\begin{center}
\begin{tabular}{r|r|r|r|r|r|r}
\hline\hline
surface &  $E_c$ & $N_l$  & $N_k$ & slab &
\multicolumn{2}{c}{$E_{a}(x)$ [eV]} \\
cell & [Ryd]  & & & type & $x=11.7^0$ & $x=17.4^0$ \\
\hline
2x1 & 30 & 6 &  32 & s & 0.39 & 0.67 \\
2x1 & 30 & 8 &  32 & s &0.38 & -- \\
2x1 & 40 & 6 &  32 & s &0.46 & --   \\
2x1 & 30 & 6 & 32 & p & 0.31 & 0.62 \\
2x1 & 30 & 6 &128 & p & 0.34 & --   \\
c(4x2) & 30 & 6 & 16 & p & 0.37 & -- \\
\hline\hline
\end{tabular}
\end{center}
\caption{
Convergence test for the adsorption barrier height on a Si dimer
tilted by 11.7 and 17.4 degrees, for different sizes of the unit cell,
cutoff-energies $E_c$,
numbers of layers in the slab $N_l$, and k-points in the surface
Brillouin zone $N_k$. On the slabs labeled ``s'' the adsorption was
modeled symmetrically on both sides, for the slabs labeled ``p'' one
side was passivated by fixed hydrogen atoms.
The c(4x2) unit cell contained two Si dimers with opposite buckling
angles, but the same transition state configuration as in the 2x1 cell
was used.}
\end{table}

\newpage
\begin{table}[h] \label{nmodes}
\begin{center}
\begin{tabular}{l|r|r}
\hline\hline
Normal mode  & adiab. TS           & sym. TS \\
             & $\hbar\omega$ [meV] & $\hbar\omega$ [meV]\\
\hline
Reaction coordinate & {\it i} 130 &{\it i} 190 \\
H--H stretch         & 180  & 260 \\
Si--H$_2$ stretch    & 100  & -- \\
Hindered cartwheel  & 165 & 115 \\
Hindered helicopter &  95 & 80 \\
Translation parallel to dimer & -- & 15  \\
Translation normal to dimer & 45 & 5 \\
\hline
Sum of real modes & 585 & 475 \\
\hline\hline
\end{tabular}
\end{center}
\caption{
The normal modes of the H$_2$ molecule at the
adiabatic and the symmetric transition state. The labels are chosen as
a hint to what the mode resembles most closely.
The missing entry is the main component of the reaction coordinate.}
\end{table}

\newpage

\begin{figure}
\caption{The calculated minimum energy path for an H$_2$ molecule
dissociating over a Si(100) surface. The trajectories of the H and
surface Si atoms are shown and the transition state is marked. The
initial, transition, and final state configurations are also indicated
as  full, dashed and long dashed
circles, respectively.}
\label{adiabat}
\end{figure}

\begin{figure}
\caption{Potential energy for an H$_2$ molecule approaching the
  Si(100) surface along the minimum energy reaction path. The full line
  assumes that the surface dimer is already in the most favorable
  configuration for sticking before the approach, whereas the dashed
  line is for the Si dimer fixed in its equilibrium configuration.
  The levels shown
  in the plot indicate the zero point energy contributions at the initial,
  transition and final state.}
\label{energy}
\end{figure}

\begin{figure}
\caption{The geometries of the adiabatic transition state (a) and the
  symmetric transition state (b). }
\label{trasta}
\end{figure}

\begin{figure}
\caption{Charge densities projected onto a plane perpendicular to the
  clean Si(100) surface which contains the surface Si dimer. From top
  to bottom the symmetric Si dimer, the most favorable Si dimer configuration
  for H$_2$ adsorption, and the equilibrium configuration are
  shown. The hydrogen molecule will most favorably adsorb on the
  charge depleted lower end of the Si dimer.}
\label{chd}
\end{figure}

\begin{figure}
\caption{Surface band structure for the Si dimer in its equilibrium
  configuration (a) and in the configuration of the adiabatic
  transition state (b). The first two energy levels above and below
  the Fermi energy (which is set to 0) are shown. Full lines refer to the
  clean surface and dotted lines to a situation with H$_2$ present
  at the transition state. In (b) the surface conduction band becomes
  depleted upon adsorption. This effect tends to lower the adsorption
  barrier. }
\label{surfbands}
\end{figure}

\begin{figure}
\caption{The total energy of the transition state $E_{TS}$, of the
clean surface $E_s$ and the activation barrier for dissociation
$E_a$ shown as a function of the buckling angle of the Si surface
dimer. While a nearly symmetric transition state is favored for small
buckling angle, the asymmetric transition state is prefered at higher
buckling angles.
The curves are fits based on the data points and the known location of the
minima. A second order polynomial in $x$ was used for $E_{TS}$ and a
third order polynomial in $x^2$ for $E_s$.}
\label{barriers}
\end{figure}

\begin{figure}
\caption{The total energy of the symmetric transition state $E_{TS}$, of the
clean surface $E_s$ and the activation barrier for dissociation
$E_a$ shown as a function of the Si dimer bond length. The lines are
fits to guide the eye. The data points
on the very right correspond to the unpaired, unrelaxed Si(100) surface.
}
\label{stretch}
\end{figure}


\begin{figure}
\caption{Entrance channel of the potential energy surface
  for an H$_2$ molecule impinging
  under $63^o$ relative to the surface normal onto
  the lower silicon atom of the surface dimer, which is frozen in the position
  of the lowest barrier. The molecule is oriented with its axis almost
  parallel to the surface. The center-of-mass distance refers to the
  midpoint between the two hydrogen atoms in the mono-hydride phase.
  The energy is given relative to the energy of
  the frozen surface plus the energy of the free H$_2$ molecule, the
  contour spacing being 50 meV. }
\label{PES}
\end{figure}

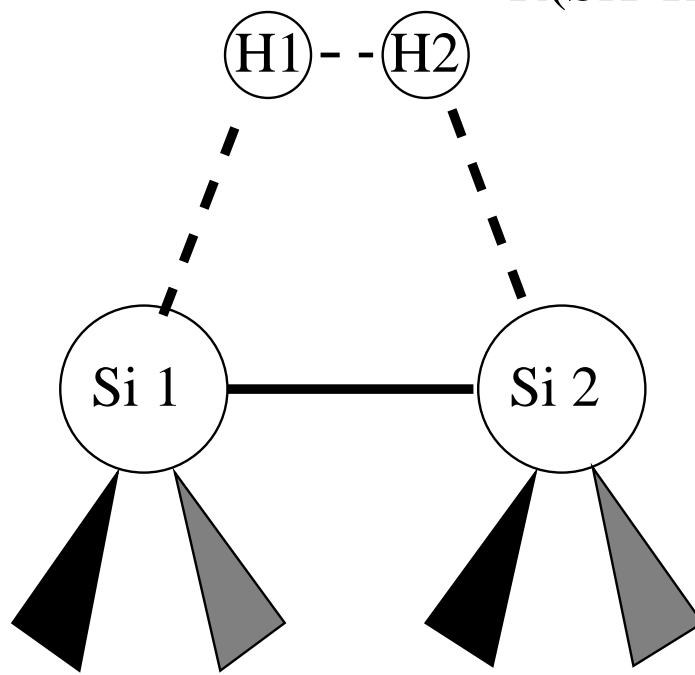
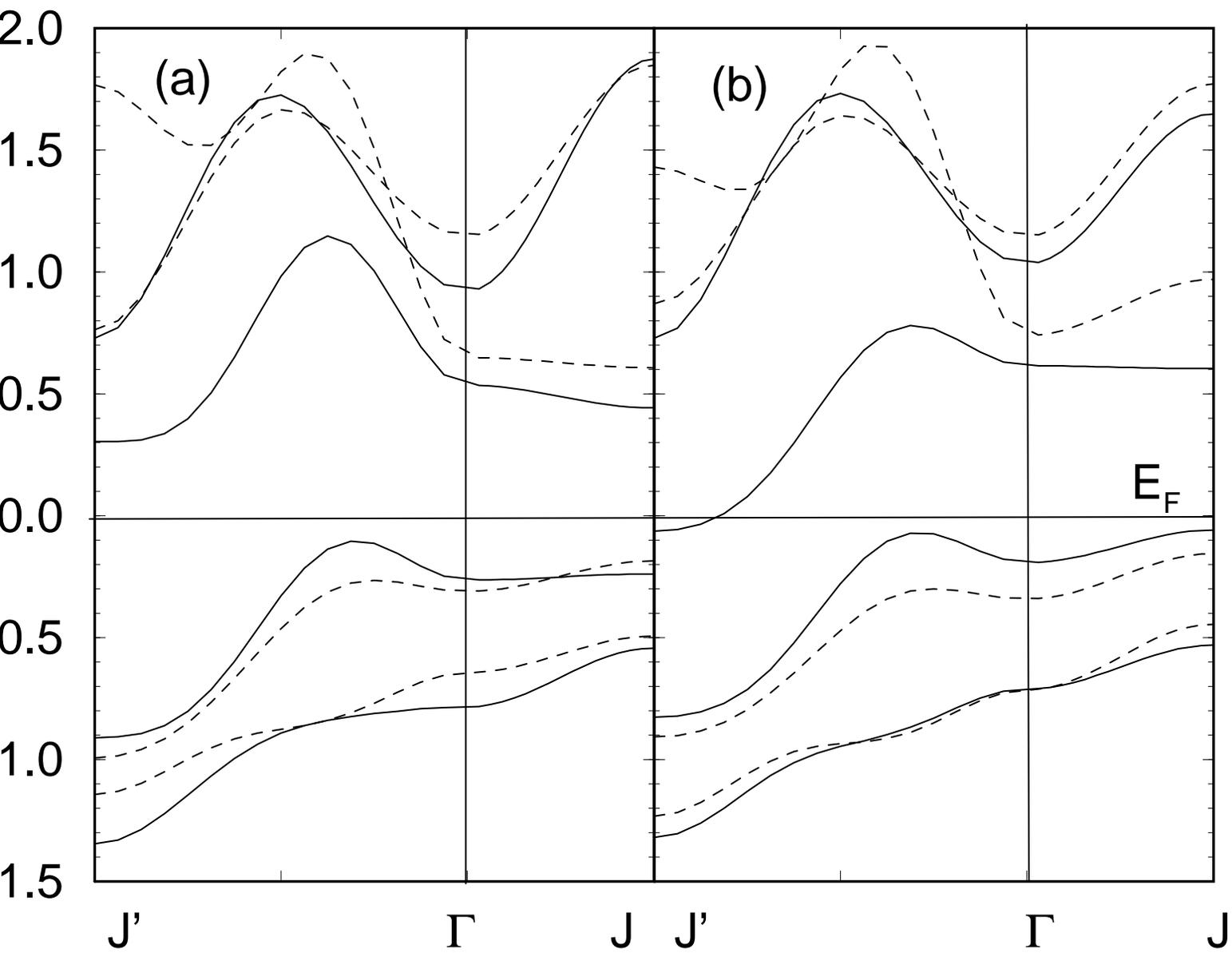
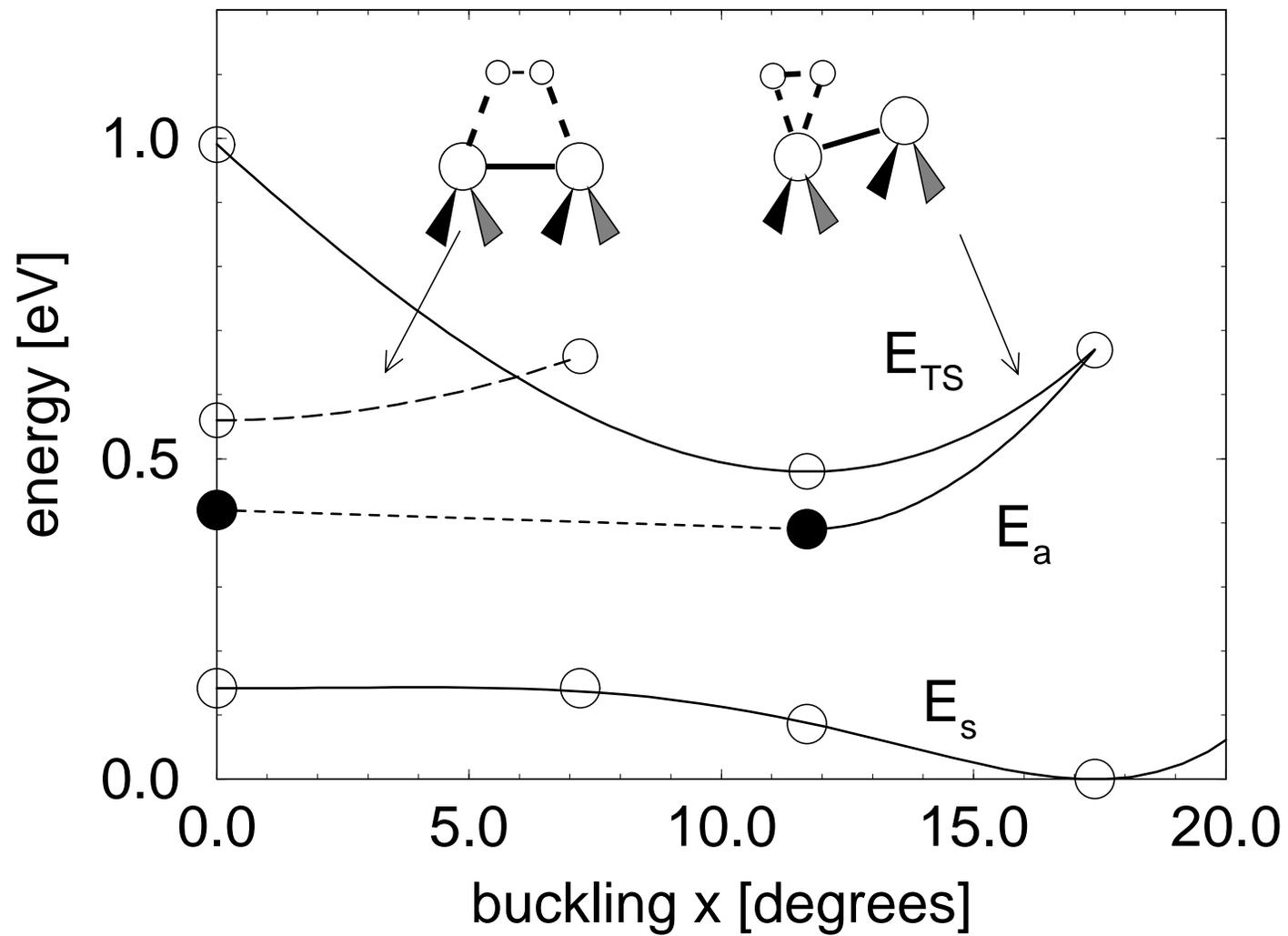
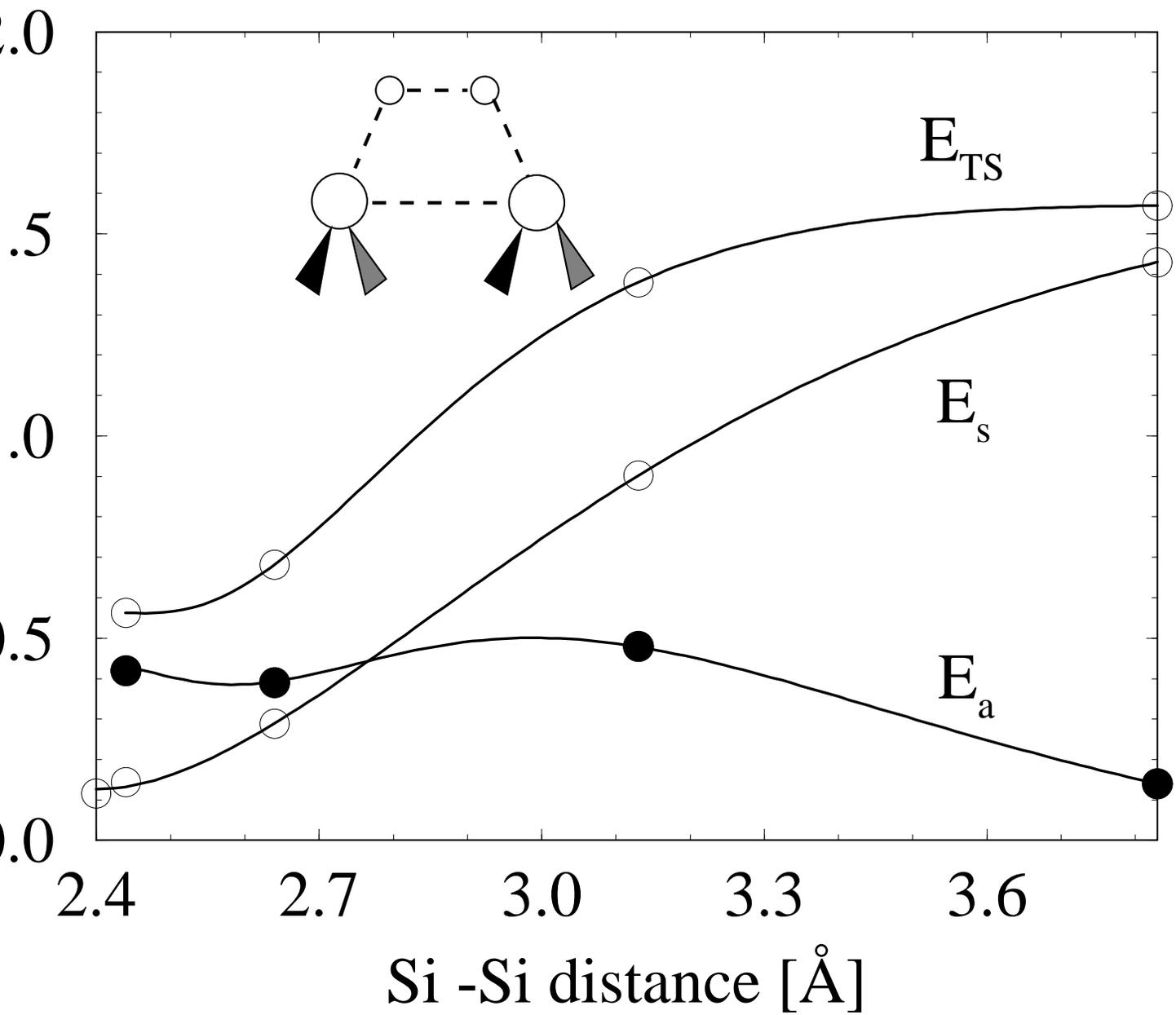
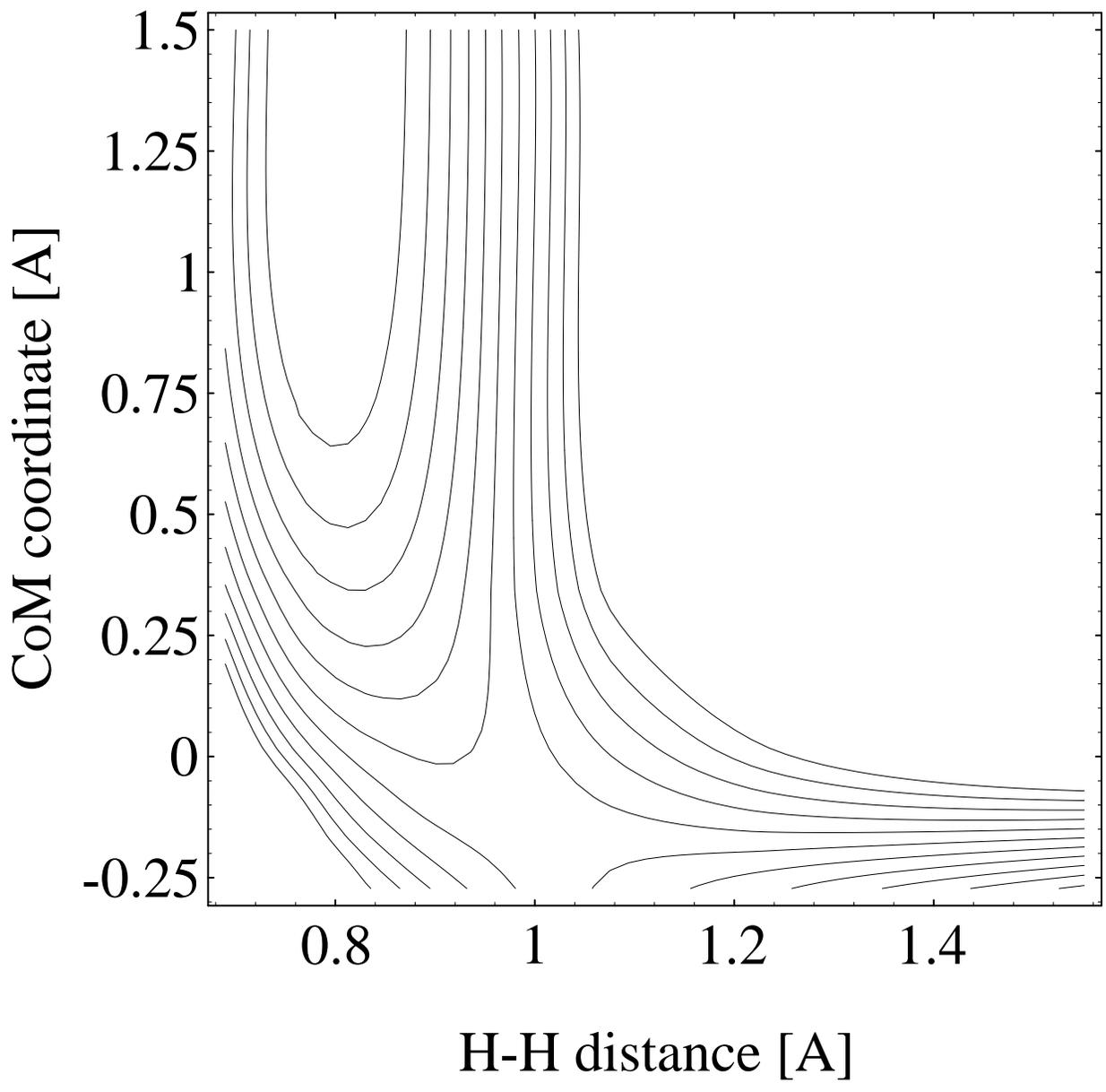
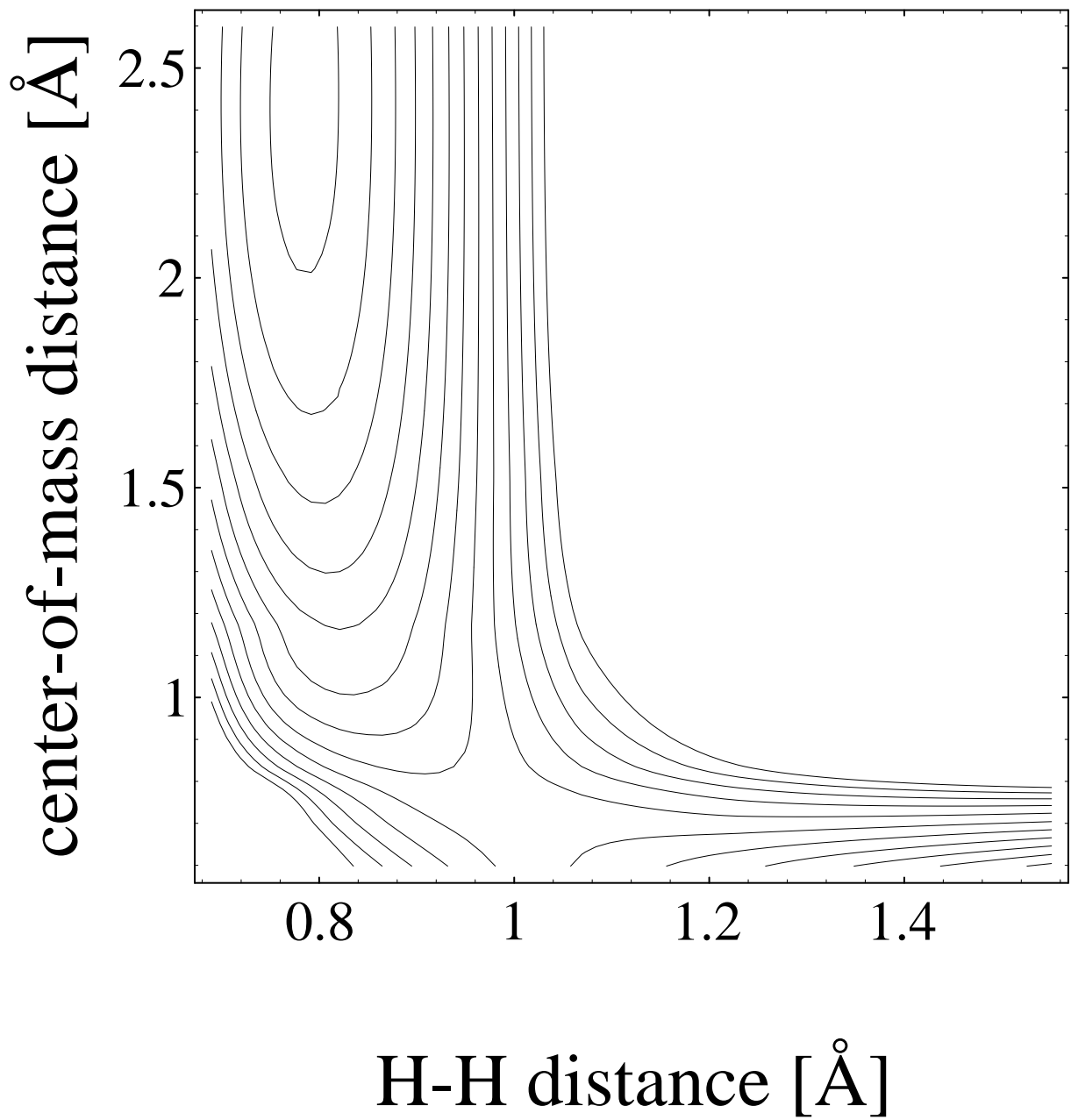
\end{document}